\documentclass[12pt]{article}
\pdfoutput=1

\usepackage{amsmath,amsfonts,amssymb}
\usepackage[nosort]{cite}
\usepackage{hyperref}
\usepackage{graphicx}
\usepackage[usenames,dvipsnames]{xcolor}
\usepackage{epsfig}
\usepackage{psfrag}	
\usepackage{tikz}
\usepackage{slashed}
\usepackage{ulem}
\usepackage[font=small,labelfont=bf,width=1.1\textwidth]{caption}
\usepackage{comment}
\usepackage[all]{xy}
\usepackage{subcaption}			

\numberwithin{equation}{section}
\usepackage{color}
\newcommand \be{\begin{eqnarray}}
\newcommand \ee{\end{eqnarray}}
\newcommand{\bea}{\begin{eqnarray}}
\newcommand{\eea}{\end{eqnarray}}
\newcommand{\beq}{\begin{equation}}
\newcommand{\eeq}{\end{equation}}
\newcommand{\bal}{\begin{equation}\begin{aligned}}
\newcommand{\eal}{\end{aligned} \end{equation}}
\newcommand{\vev}[1]{{\left< {#1} \right>}}

\DeclareMathOperator{\tr}{tr}

\DeclareMathOperator{\vol}{vol}

\newcommand{\cN}{{\mathcal N}}
\newcommand{\cO}{{\mathcal O}}

\newcommand{\so}{\mathfrak{s}\mathfrak{o}}

\newcommand{\im}{i}

\newcommand{\diff}{d}
\newcommand{\fin}{\text{finite}}
\newcommand{\RicciScalar}{R}
\newcommand{\RicciTensor}{R}
\newcommand{\cAnomaly}{\left( \partial n \right)^2}
\usepackage{romanbar} 
\newcommand{\IIFundForm}{\Romanbar{2}}
\usepackage{tensor} 
\newcommand{\norm}[1]{\left| #1 \right|}
\newcommand{\ho}[2][]{\mathcal{O}(#2^{#1})}
\newcommand{\RiemannTensor}{R}

\newcommand{\address}[1]{\vbox{\center\em#1}}
\renewcommand{\title}[1]{\vbox{\center\huge{#1}}\vspace{5mm}}

\begin{document}

\begin{titlepage}
\begin{center}

\vspace*{20mm}

\renewcommand{\thefootnote}{$\alph{footnote}$}
\vspace{10mm}

\title{Surface operators in the 6d $\mathcal{N} = (2,0)$ theory}
Nadav Drukker,\footnote{\href{mailto:nadav.drukker@gmail.com}{nadav.drukker@gmail.com}}
Malte Probst,\footnote{\href{mailto:mltprbst@gmail.com}{mltprbst@gmail.com}}
and 
Maxime Tr\'epanier\footnote{\href{mailto:trepanier.maxime@gmail.com}{trepanier.maxime@gmail.com}}
\address{Department of Mathematics, King's College London,\\The Strand, London WC2R
2LS, United Kingdom}

\renewcommand{\thefootnote}{\arabic{footnote}}
\setcounter{footnote}{0}

\end{center}

\vspace{8mm}
\abstract{
\normalsize{
\noindent
The 6d $\cN=(2,0)$ theory has natural surface operator observables, which are akin in many 
ways to Wilson loops in gauge theories. We propose a definition of a ``locally BPS'' 
surface operator and study its conformal anomalies, the analog of the conformal dimension 
of local operators. We study the abelian theory and the holographic dual of the 
large $N$ theory refining previously used techniques. 
Introducing non-constant couplings to the scalar fields allows for 
an extra anomaly coefficient, which we find in both cases to be related to one of the 
geometrical anomaly coefficients, suggesting a general relation due to supersymmetry. 
We also comment on surfaces with conical singularities.}}
\vfill

\end{titlepage}

\section{Introduction}
\label{sec:intro}

Understanding the six dimensional $\mathcal{N} = (2,0)$ superconformal field theory 
is one of the most intriguing problems in theoretical physics. In this paper we revisit 
the most natural observables in this theory, surface operators~\cite{Howe:1997ue}.
If we define the theory as arising from $N$ coincident M5-branes, the simplest
surface operators correspond to the endpoints of M2-branes \cite{Strominger:1995ac}.

In some ways the surface operators in six dimensions are analogous to Wilson loops in lower 
dimensional gauge theories. Wilson loops are the boundaries of fundamental strings, which are 
the dimensional reduction of M2-branes, and indeed one obtains Wilson loops in compactifications 
of the 6d theory with surface operators. Wilson loops are not only interesting due to their 
physical importance, they are also accessible to many perturbative and non-perturbative 
calculational tools in supersymmetric field 
theories: Feynman diagrams, holographic descriptions~\cite{Rey:1998ik, maldacena:1998im, 
drukker:1999zq}, localization~\cite{pestun:2007rz}, the defect CFT framework and 
associated OPE techniques~\cite{cooke:2017qgm, Giombi:2017cqn}, integrability~\cite{Drukker:2012de, 
Correa:2012hh}, duality to scattering amplitudes~\cite{Alday:2007hr} and more.
See for instance a recent survey of these techniques, as applied to supersymmetric 
Wilson loops in ABJM theory~\cite{Drukker:2019bev}.

We do not expect all these techniques to extend to surface operators in six dimensions, 
but it is worthwhile to examine which of them may work, and we hope that some 
calculations may lead to exact results applicable for all $N$. Here 
we take the first step in such an examination, defining the notion of 
a ``locally BPS surface operator'' and studying basic properties 
of their anomalies. This is mainly based on previous work~\cite{graham:1999pm, 
henningson:1999xi, hen99, gus03, gustavsson:2004gj}, which we modify and refine in several ways.

As reviewed in the next section, the evaluation of generic surface operators 
leads to logarithmic divergences. The anomaly depends on the geometry of the 
surface, as well as intrinsic properties of the operator which are captured by 
three numbers, known as \emph{anomaly coefficients}~\cite{deser:1993yx}. 

The ``locally BPS'' operator couples to the scalar fields via a unit 
5-vector $n^i$. This can be viewed as a coupling to an R-symmetry background, 
and for non-constant $n^i$ we find a new anomaly, proportional to $(\partial n)^2$, with its 
own anomaly coefficient.

We perform explicit calculations of the three geometrical and one background 
coefficients in both the free theory at $N=1$ and the holographic description 
valid at large $N$. An examination of our results reveals that the new anomaly 
coefficient matches (up to a sign) one of the geometric ones in both regimes. 
We present here a simple argument, relying on supersymmetry, why we expect 
this relation to hold for all $N$. 
A more rigorous proof of this relation based 
on the application of defect CFT techniques to surface operators 
will be presented in \cite{Drukker:2020atp}.

Beyond the study of $\mathcal{N}=(2,0)$ superconformal symmetry, surface operators in 
conformal field theories  have drawn interest within a number of different contexts. Recent 
work on entangling surfaces in 4d~\cite{solodukhin:2008dh,myers:2012vs,bianchi:2015liz,Jensen:2018rxu} 
and theories with boundaries~\cite{Fursaev:2015wpa, Solodukhin:2015eca, Herzog:2017xha} uses some techniques
which apply in our case as well. In particular, the classification of local conformal invariants of
surfaces is independent of the codimension and translates to the 6d case \cite{Schwimmer:2008yh}. 

Surface operators in the $\mathcal{N} = (2,0)$ theory have been studied both from 
a field theory perspective~\cite{henningson:1999xi, hen99, 
gus03,gustavsson:2004gj} and using holography~\cite{berenstein:1998ij,
graham:1999pm}. Corresponding soliton solutions of the M5-brane equations of motion 
have been discussed in the literature under the moniker of self-dual strings~\cite{Howe:1997ue}.

The resemblance to Wilson loops is evident in both the field theoretic and the holographic 
approach. In the former, for $N=1$ as is studied in Section~\ref{sec:abel}, 
we define the surface operator in analogy to the Maldacena-Wilson 
loops~\cite{maldacena:1998im} as
\beq
\label{WilsonSurface_Definition}
V_\Sigma = \exp \int_\Sigma \left( \im B^+ - n^i \Phi_i \vol_\Sigma \right),
\eeq
where $B^+$ is the pullback of the chiral 2-form to the surface $\Sigma$ and 
$\Phi^i$ are the scalar fields.

Since for $N>1$ there is no realisation of the theory in terms of fundamental fields, 
we cannot give an analogous definition of the surface operator. However, by analogy with 
Wilson loops \cite{Rey:1998ik, maldacena:1998im, drukker:1999zq}, in the large $N$ limit, 
these operators in the fundamental representation have a nice holographic 
dual as M2-branes ending on the surface and extending into 
the $AdS_7 \times S^4$ bulk, as discussed in Section~\ref{sec:hol}.
In the absence of a scalar coupling breaking the $\so(5)$ R-symmetry, these would
be delocalised on the $S^4$ \cite{Alday:2007he, Polchinski:2011im}. At leading
order, we need only consider minimal 3-volumes~\cite{maldacena:1998im,
graham:1999pm} (similar to the minimal surfaces of interest in the Wilson loop
case~\cite{Rey:1998ik, maldacena:1998im, drukker:1999zq}), and to find the anomaly, which is a
local quantity, it is enough to understand the volume close to the $AdS$
boundary.  High-rank (anti-)symmetric representations are dual to configurations
involving M5 branes shrinking to the surface on the boundary of $AdS_7$ and have
been considered in \cite{Chen:2007ir, chen:2008ds, mori:2014tca,
Rodgers:2018mvq}.

The definition in \eqref{WilsonSurface_Definition} includes BPS operators. Simple 
examples are the plane or sphere with constant unit $n^i$. Other examples are briefly 
discussed in Section~\ref{sec:abel} and will be explored in more detail 
elsewhere~\cite{S3-surfaces}. We call operators with generic $\Sigma$ and unit length $n^i$ 
``locally BPS'', and show that they possess some nice properties, in particular 
that all power law divergences cancel.

In the next section we recall the structure of surface operator anomalies 
and introduce the anomaly coefficients. We evaluate these anomaly coefficients for 
the two known realisations of
the $\mathcal{N} = (2,0)$ theory; first as the theory of a single M5-brane
($N=1$)~\cite{claus:1997cq}, for which the equations of motion are known~\cite{howe:1997fb}, and
second, using holography (for the large $N$ limit) from M-theory on the 
${AdS}_7 \times S^4$ background~\cite{maldacena:1997re} found in~\cite{Freund:1980xh}. The resulting anomaly coefficients 
are presented in equations~\eqref{eqn:FTanomalycoeff} and~\eqref{eqn:GWanomalycoeff}.
After performing the free field and holographic calculations, we address in 
Section~\ref{sec:singularity} surfaces with singularities.
We discuss our results in Section~\ref{sec:conclusion} and offer a simple 
argument for the relation between two of the anomaly coefficients.
We collect some technical tools in appendices. Our conventions can
be found in Appendix~\ref{sec:convention}. Details of the geometry of
submanifolds are compiled in Appendix~\ref{sec:GeometrySubmanifold}.
Appendix~\ref{sec:GeodesicPointSplitting} contains an alternative, more
geometric derivation of the field theory results in Section~\ref{sec:abel}.

\section{Surface anomalies}
\label{sec:anomaly}

The most natural quantities associated to surface operators in conformal field 
theories are their anomaly coefficients. To understand their origin, 
note that, unlike line operators, the expectation values of surface
operators typically suffer from ultraviolet divergences, which cannot be removed 
by the addition of local counterterms. The regularised expectation value satisfies
\beq
  \log{\vev{V_\Sigma}} \sim
  \log{\epsilon} \int_\Sigma \vol_\Sigma \mathcal{A}_\Sigma  +
  \fin,
\eeq
where $\epsilon$ is a regulator, $\mathcal{A}_\Sigma$ is known as the 
\emph{anomaly density}, and we suppressed possible power-law divergences. 

$\mathcal{A}_\Sigma$ is scheme independent and 
indicates an anomalous Weyl symmetry, since for a constant
rescaling $g \to e^{2 \omega} g$, the expectation value
varies as
\beq
  \log{\vev{V_\Sigma}_{e^{2 \omega}g}} - \log{\vev{V_\Sigma}_{g}} = \omega
  \int_\Sigma \vol_\Sigma \mathcal{A}_\Sigma\,,
\eeq
where the subscript $\vev{\bullet}_g$ denotes the background metric.

The anomaly is constrained by the Wess-Zumino consistency 
condition~\cite{deser:1993yx,Boulanger:2007st} to be
conformally invariant.  In dimensions $d \ge 3$, the local
geometric conformal invariants for a 2d submanifold, which have been
classified in~\cite{Schwimmer:2008yh}, are
\begin{itemize}
\item[]
\makebox[2.3cm][r]{$\RicciScalar^{\Sigma}$:} The Ricci scalar of the induced metric $h_{ab}$ on $\Sigma$.
\item[]
\makebox[2.3cm][r]{$H^2 + 4 \tr{P}$:} 
$H^\mu$ is the mean curvature, $P_{ab}$ the  
pullback of the Schouten tensor \eqref{eqn:Schouten}.
\item[]
\makebox[2.3cm][r]{$\tr{W}$:}
$W_{abcd}$ is the pullback of the Weyl tensor.
\end{itemize}
Under conformal transformations, the first two change by a total
derivative (type A anomalies) and the last is itself conformally invariant (type B).

As we allow for variable couplings to the scalars, parametrised by a unit 5-vector $n^i$, 
we find an extra potential type B Weyl anomaly associated to it:
\begin{itemize}
\item[]
\makebox[2.3cm][r]{
$(\partial n)^2$} $\equiv\partial^a n^i\partial_a n_i$.
\end{itemize}
This is (up to total derivatives) the only quantity of the correct dimension that can be 
constructed using only $n$. 

The anomaly of a surface operator in any 6d ${\cal N}=(2,0)$ theory then takes the form 
\beq
  \mathcal{A}_{\Sigma}
  = \frac{1}{4\pi} \left[
  a_1 \mathcal{R}^{\Sigma}
  + a_2 \left( H^2 + 4 \tr{P} \right)
  + b \tr{W}
  + c \cAnomaly \right].
  \label{eqn:SCAnomaly}
\eeq
The anomaly coefficients $a_1$, $a_2$, $b$ and $c$ depend on the theory (that is on $N$) 
and the type of surface operator (which, at least at large $N$, is specified by the representation 
of the $A_{N-1}$ algebra~\cite{DHoker:2008rje, bachas:2013vza}), but not on its geometry 
or $n$. They are the focus of this paper.

Let us mention that there exists another commonly used basis where
\beq
\mathcal{A}_\Sigma = \frac{1}{4 \pi} \left[
  a \RicciScalar^{\Sigma} + b_1 \tr{\tilde{\IIFundForm}^2} + b_2 \tr{W} +c(\partial n)^2
  \right],
\eeq
where $\tilde{\IIFundForm}^\mu_{ab}$ is the traceless part of the second fundamental form (see~\eqref{eqn:tracelessIIFundForm}). These bases are related through the
Gauss-Codazzi equation~\eqref{eqn:GaussCodazzi}. The relation between the
coefficients is then
\bal
  a_1 &= -b_1 + a \,,\qquad& 2 a_2 &= b_1 \,,\qquad& b &= b_2 + b_1\,,\\
  a &= a_1 + 2 a_2\,,\qquad& b_1 &= 2 a_2 \,,\qquad& b_2 &= b - 2 a_2\,.
\eal

Some results about these anomaly coefficients are known for surface
defects in generic CFTs. The bound $b_1 < 0$ was derived in~\cite{bianchi:2015liz} by showing that
$b_1$ captures the 2-point function of the displacement operator, which is
positive by unitarity. Similarly, it was shown
in~\cite{Lewkowycz:2014jia,bianchi:2015liz} that $b_2$ is calculated by the
one-point function of the stress tensor in the presence of the surface defect
(this was also conjectured in~\cite{Drukker:2008wr}). Assuming that the average
null energy condition holds in the presence of defects also leads to a bound
$b_2 > 0$~\cite{Jensen:2018rxu}. 

For the surface operators at hand, these anomaly coefficients were also calculated
previously. At large $N$, the first such result was a calculation of the 1/2-BPS
sphere~\cite{berenstein:1998ij}, with total anomaly $-4N$, implying $a_1^{(N)} +
2 a_2^{(N)} = -2N$, at leading order at large $N$. This was soon followed by the
more detailed result $a_2^{(N)} = - N$ and $a_1^{(N)} = b^{(N)} = 0$~\cite{graham:1999pm}.

More recently, it was conjectured that $\cN=(2,0)$ supersymmetry imposes $b = 0$
(or $b_1 = -b_2$) for any $N$~\cite{Bianchi:2019sxz}. $a$ and $b_2$ were
calculated at any $N>1$ (and for any representation) by studying the holographic
entanglement entropy in the presence of surface
operators~\cite{Gentle:2015jma,Rodgers:2018mvq,Jensen:2018rxu,Estes:2018tnu}.
This result is also supported by a recent calculation based on the
superconformal index~\cite{Chalabi:2020iie}, which suggests that it is exact.

The anomaly coefficient $c$ has previously not been discussed, to our knowledge.

\section{Abelian theory with $N=1$}
\label{sec:abel}

In this section we study the anomaly coefficients of the surface operator in the 
abelian $(2,0)$ theory.
This is the theory of a single M5-brane and the degrees of freedom form the 
tensor supermultiplet of the $\mathfrak{osp}(8^*|4)$ symmetry algebra.
It consists of three fields~\cite{gunaydin:1984wc,gunaydin:1999ci} 
(see~\cite{Buican:2016hpb,cordova:2016emh} for an overview 
of superconformal multiplets in various dimensions)
\begin{itemize}
\item
A real closed self-dual 3-form $H = \diff B^+$.
\item
A chiral spinor $\psi_{\alpha \check{\alpha}}$ subject to the symplectic
Majorana condition $\bar{\psi} = -c \Omega \psi$~\eqref{eqn:symplecticMajorana}
where $c$ and $\Omega$ are charge conjugation matrices, see~\eqref{eqn:gammarep}.
\item
Five real scalar fields $\Phi^i$.
\end{itemize}
These fields
transform into each other under superconformal transformations as
\cite{claus:1997cq}
\bal
\label{eqn:susy}
    \delta_\varepsilon B^+_{\mu\nu} &= \varepsilon(x) \gamma_{\mu\nu} \psi\,,\\
    \delta_\varepsilon \psi &=
    -\gamma^\mu \partial_\mu \Phi^i
    \check{\gamma}_i \bar{\varepsilon}(x)
    + \frac{1}{12} \gamma^{\mu\nu\rho}
    H_{\mu\nu\rho} \bar{\varepsilon}(x)
    + 4 \Phi^i \check{\gamma}_i \varepsilon^1 \,,\\
    \delta_\varepsilon \Phi^i &= - \varepsilon(x) \check{\gamma}^i \psi\,.
\eal
The parameter $\bar{\varepsilon}(x)$ is an antichiral
spinor of the form
$\bar{\varepsilon}_{\dot{\alpha} \check{\alpha}}(x) =
\bar{\varepsilon}^0_{\dot{\alpha} \check{\alpha}} +
\tensor{(\bar{\gamma}_\mu)}{_{\dot{\alpha}}^\beta} x^\mu \varepsilon^1_{\beta
\check{\alpha}}$, where $\bar{\varepsilon}^0$ and $\varepsilon^1$ are constant
spinors parametrising, respectively, the supersymmetry and special supersymmetry
transformations. Our spinor conventions are summarised in
Appendix~\ref{sec:convention}.

\subsection{Surface operators and BPS condition}
\label{sec:Projector}

We define the surface operators $V_\Sigma$ of the abelian theory as
in~\eqref{WilsonSurface_Definition}.  We restrict to space-like surfaces 
in flat 6d Minkowski space (with mostly positive signature). Null surfaces could
be interesting by analogy with null polygonal Wilson loops, which are dual to
scattering amplitudes in $\cN = 4$ SYM~\cite{Alday:2007hr}, but lie beyond
the scope of this work (see however~\cite{Bhattacharya:2016ydi}).

A surface operator is BPS provided that its
variation under the supersymmetry transformations~\eqref{eqn:susy} vanishes
\beq
  \delta_\varepsilon V_\Sigma
  = -\int \varepsilon(x) \left[ \frac{\im}{2} \gamma_{\mu\nu} \partial_a x^\mu
    \partial_b x^\nu \epsilon^{ab}
  - n^i \check{\gamma}_i \right] \psi \vol_\Sigma V_\Sigma = 0\,.
  \label{eqn:BPSvar}
\eeq
Since this is an integral over the insertion of an operator $\psi$ along the
surface, this is satisfied only when the integrand vanishes at every point along the
surface, leading to the projector equation
\beq
\varepsilon\Pi_-=0\,,
\qquad
\Pi_-=\frac{1}{2} - \frac{\im}{4}\partial_a x^\mu \partial_b x^\nu \epsilon^{ab}
\frac{n^i}{n^2} \gamma_{\mu\nu}\check{\gamma}_i\,.
\label{eqn:BPSXproj}
\eeq
If we impose that $n^2 \equiv n^i n^i = 1$, then $\Pi_-$ is a half rank projector and
otherwise it is a full rank matrix.  In the case of a planar surface with 
constant unit $n^i$, this is a
single condition, so the surface preserves 16
supercharges, i.e. is $1/2$-BPS.%
\footnote{%
The BPS condition for a surface operator extended in the time-like direction can
be obtained by Wick-rotation to
$$
  V_{\Sigma}^\text{timelike} = \exp{ \left[ \im \int_\Sigma B^+ - \Phi \vol_\Sigma \right]}.
$$}

In analogy to Wilson loops in 4d theories, it is natural to discuss ``locally BPS operators''~\cite{drukker:1999zq}, where 
the equations \eqref{eqn:BPSXproj} are satisfied at every point along the surface, but without 
a global solution. This amounts to the requirement $n^2=1$, and as shown below, leads to the 
cancellation of all power-like divergences in the evaluation of the surface operator.

One can also look for surfaces, other than planes, that preserve some smaller fraction 
of the supersymmetry by relating $n^i(\sigma)$ to $x^\mu(\sigma)$ and its derivatives. One 
simple way to realise this is for surfaces with the geometry ${\mathbb R}\times S$, for 
some curve $S\subset{\mathbb R}^{1,4}$. Upon dimensional reduction this becomes a Wilson loop 
in 5d maximally supersymmetric Yang-Mills (or 4d upon further dimension reduction). 
Then one can choose $n^i$ to follow the construction of globally BPS Wilson loops 
of~\cite{zarembo:2002an} or~\cite{drukker:2007qr} to find globally BPS surface operators. 
Indeed this was realised recently in \cite{mezei:2018url} (see also~\cite{Lee:2006gqa}).

There are further examples of globally BPS surface operators, which do not follow this 
construction. The simplest is the spherical surface, but there are several other classes of such 
operators, which will be explored elsewhere \cite{S3-surfaces}.

\subsection{Propagators}

Since the abelian theory is non-interacting, the expectation value of $V_\Sigma$ reduces to
\beq
\label{I_Sigma_Definition}
\log{\vev{V_\Sigma}} =\frac{1}{2} \int
\left[-\vev{ B^+(\sigma) B^+(\tau) }
+ \vev{\Phi_i(\sigma) \Phi_j(\tau)} n^i(\sigma)n^j(\tau)
\sqrt{h(\sigma)h(\tau)}d^2\sigma\,d^2\tau\right],
\eeq
where $h$ is the determinant of the induced metric on $\Sigma$.  Evaluating this requires
expressions for the propagators of the tensor and scalar fields. 

While one would preferably derive the propagators from an action, none is  
readily available. Many actions for the abelian $\mathcal{N} = (2,0)$ theory have been proposed
over the years, but they all suffer from some pathologies regarding the 
self-dual 2-form (see \cite{Witten:1996hc, pasti:1996vs, claus:1997cq, bagger:2012jb, 
Lambert:2019khh} for examples of available actions, and 
\cite{Sen:2019qit, Lambert:2019diy, Andriolo:2020ykk} and references therein for recent 
accounts of the various approaches in the abelian
theory).  In any case, gauge fixing and inverting the
kinetic operator is not straightforward.

\subsubsection{Tensor structure}

We sidestep these obstacles by determining the propagators in other ways. 
The scalar propagator in flat 6d is fixed by conformal symmetry to be
\beq
\vev{\Phi_i(x) \Phi_j(y)} = \frac{ C_\Phi \delta_{ij}}{\norm{x-y}^4}\,.
\eeq
The proportionality constant depends on the normalisation of the fields. It
could be determined from an action, but in its absence it is fixed by supersymmetry below.

The more complicated question is the self-dual 2-form propagator. Let us start by considering 
an unconstrained 2-form field $B$ with a free Maxwell type action
\beq
  S_\text{tot} \propto{} \int \diff^6 x\, B^{\mu \nu} 
  \left( -(\delta_\mu^\rho\delta_\nu^\sigma-\delta_\nu^\rho\delta_\mu^\sigma) \partial^2 
  + 4 (1-\alpha) \partial_{\mu} \partial^{\rho} \delta_{\nu}^{\sigma} \right) B_{\rho \sigma}\,,
\eeq
were $\alpha$ is a gauge fixing parameter. In Feynman gauge $\alpha = 1$, this gives 
the propagator
\beq
\label{eqn:TensorPropagatorCB}
\vev{B^{\mu \nu}(x)B_{\rho \sigma}(y)} = \frac{C_B (\delta_\mu^\rho\delta_\nu^\sigma-\delta_\nu^\rho\delta_\mu^\sigma)}
{\norm{x-y}^4}\,.
\eeq
Now we decompose the field into its self-dual and anti-self-dual parts 
$B_{\mu\nu}=B_{\mu\nu}^++B_{\mu\nu}^-$ and try to deduce the propagators for each component.

Since there is no covariant 4-tensor satisfying the self-duality properties of a
mixed correlator $\vev{B^+ B^-}$, we can decompose
\beq
\vev{BB} = \langle B^+ B^+\rangle + \langle B^- B^-\rangle\,.
\eeq
The two terms on the right hand side need not be identical, but the difference between them 
should be parity-odd.%
\footnote{The two-dimensional analogue is instructive. 
The propagator of a free boson in complex coordinate $z$ is given by
$$
\vev{\phi(z) \phi(0)} = \log |z|^2\,,
$$
while for a (anti-)chiral boson one finds 
$$
\vev{\phi_+(z) \phi_+(0)}= \log z\,,
\qquad
\vev{\phi_-(z) \phi_-(0)}= \log \bar z\,.
$$
Indeed the sum reproduces the free boson propagator, but the two differ by 
a parity-violating imaginary part.} 
The only such term of the right scaling dimension which we can write down is
\beq
\langle B^+ B^+\rangle - \langle B^- B^-\rangle \propto \epsilon\indices{^\mu ^\nu _\rho _\sigma _\kappa _\lambda} 
\frac{x^\kappa y^\lambda}{\norm{x-y}^6}.
\label{eqn:parity}
\eeq
However, terms of this type do not contribute to \eqref{I_Sigma_Definition}, since 
the integration is symmetric in $x$ and $y$. Therefore, for the purpose of our
calculation we can take $\vev{B^+ B^+}= \vev{B B}/2$. Note that in curved space 
we can add to the right hand side a term proportional to the Weyl tensor 
with all the required symmetries.

\subsubsection{Normalisation}

The normalisation of the tensor field propagator is fixed by the assumption that 
the surface operator defined in~\eqref{WilsonSurface_Definition} corresponds to 
a single unit of quantised charge. First, for any closed surface $\Sigma$, we can 
rewrite the surface operator (without scalars) in terms of the field strength as
\begin{equation}
\exp \int_\Sigma \im B^+ = \exp \int_V \im H,
\end{equation}
where $\partial V=\Sigma$. 
In order for this to be well-defined, any two such $V$ with the same 
boundary must yield the same result. Equivalently, for every closed 3-manifold $V$
\begin{equation}
\int_V H \in 2\pi \mathbb{Z}\,,
\end{equation}
and similarly for $\ast H$.

Now consider a flat surface operator in the $(x^1, x^2)$ plane, which we view as
a source for the self-dual $B$ field. The solution to the equations of
motion would be given by convoluting the propagator with this source. Using the
expression in~\eqref{eqn:TensorPropagatorCB} and adding the factor $1/2$ to
account for restricting to the self-dual sector, we get
\begin{equation}
\label{eqn:convolution}
B_{\mu\nu}(x) = \int_{{\mathbb R}^2} 
\frac{1}{2} \frac{C_B(\delta^1_\mu\delta^2_\nu-\delta^1_\nu\delta^2_\mu)}{|x-y|^4}
\,\diff y^1\,\diff y^2\,.
\end{equation}
Again, because we don't know the self-dual propagator, the field strength we
obtain is not self-dual, but the quantisation condition should still be
satisfied. Imposing that the charge enclosed in a transverse sphere is quantised
leads to
\beq
  \int_{S^3} \ast H = 2 \pi^3 C_B = 2 \pi \qquad
  \Rightarrow \qquad
  C_B = \frac{1}{\pi^2}\,.
\eeq

The normalisation of the scalar propagator is then fixed by supersymmetry. A simple way to 
implement that is to compare with the classical BPS solution of the self-dual
string~\cite{Howe:1997ue} which gives%
\footnote{The absence of power-law divergences in the calculation 
in the next section is also a hint that this is indeed the correct proportionality.}
$2 C_\Phi = C_B$. Overall, we are left with 
\begin{subequations}
\label{eqn:Propagators}
\begin{align}
  \label{Propagator_Phi}
  \vev{\Phi_i(x) \Phi_j(y)} &= \frac{\delta_{ij}}{2\pi^2 \norm{x-y}^4}\,, \\
  \vev{B^+_{\mu \nu}(x) B^+_{\rho \sigma}(y)} 
  &= \frac{\delta_{\mu\rho}\delta_{\nu\sigma}-\delta_{\mu\sigma}\delta_{\nu\rho}}
  {2\pi^2 \norm{x-y}^4}\,.
\label{Propagator_B}
\end{align}
\end{subequations}

We emphasise that this normalisation is obtained by imposing a quantisation condition 
on the self-dual sector of an unconstrained $B$-field. 
This follows the discussion in \cite{Witten:1996hc}, however some 
caution is warranted. A proper treatment of the quantisation of a self-dual
two-form could add a factor of 2 on the right hand side
of~\eqref{eqn:convolution}, changing the resulting anomaly coefficients.

With the flat space propagators we are able to determine the anomaly
coefficients $a_1$, $a_2$, and $c$. The calculation of $b$, however, requires
the curved space propagator, where the 
right-hand side of~\eqref{eqn:parity} could pick up contributions whose integral 
does not vanish. Since we do not know how to fix these terms, we cannot 
determine $b$.

Note though that we can calculate the contribution of the scalars to the anomaly 
coefficient $b$. The propagator of a conformal scalar in a curved background 
can be expanded in powers of the geodesic distance~\cite{gustavsson:2004gj}, 
and the contribution to the anomaly coefficient $b$ is read off as $-1/3$.

If we give up the requirement of self-duality, we can use the short-distance
expansion of an unconstrained 2-form propagator on curved space, which has been
computed in~\cite{henningson:1999xi, gus03}, and again, the Weyl tensor of the
background explicitly contributes to the curvature corrections. Halving that to
try to account for self-duality and adding to it the contribution from the
scalars, one obtains $b = - 4/3$~\cite{gustavsson:2004gj}. This is in
disagreement with the conjecture $b=0$~\cite{Bianchi:2019sxz} and therefore one
may not trust it.

\subsection{Evaluation of the anomaly}
\label{sec:FTanomaly}

With the propagators at hand, we can compute the expectation value of the 
surface operator by evaluating the integrals in \eqref{I_Sigma_Definition}. 
Generically, these integrals are divergent and must be regularised. 

In this section we take a rather naive approach of placing a hard UV cutoff on the 
double integral \eqref{I_Sigma_Definition}, so as to restrict $|\sigma-\tau|>\epsilon$
(where the distance is measured with the induced metric), the same regularisation 
that is used in~\cite{henningson:1999xi}. 
A different regularisation is employed in \cite{gustavsson:2004gj}, 
where the surface is assumed to be contained within a 5d linear subspace of 
${\mathbb R}^6$ and the two copies of the surface are displaced by a distance 
$\epsilon$ in the 6th direction. This restriction to ${\mathbb R}^5$ must still yield the 
correct answer, since even for surfaces in 4d the geometric invariants 
in the anomaly \eqref{eqn:SCAnomaly} are independent of each other. Still, in 
Appendix~\ref{sec:GeodesicPointSplitting} we redo the calculation removing this 
assumption by displacing the two copies of the surface along geodesics in the direction 
of an arbitrary normal vector field. That approach could be important for the calculation 
of surface operators in four dimensions, where the restriction to a 3d linear subspace 
does not allow to resolve all the anomaly coefficients.

To find the anomalies we only need the short-distance behaviour of the propagators, 
so we use normal coordinates $\eta^a$ about a point $\sigma$ on $\Sigma$. The notations 
and required geometry are presented in Appendix~\ref{sec:GeometrySubmanifold}. 

Starting from the scalar contribution to \eqref{I_Sigma_Definition}, the integrand is
\beq
\frac{1}{4\pi^2}\frac{n^i(\sigma)n^i(\tau)}{|x(\sigma)-x(\tau)|^4}\sqrt{h(\sigma)}\sqrt{h(\tau)}\,.
\label{eqn:FTscalarintegrand}
\eeq
Using $n^i n^i = 1$ and \eqref{VolumeFactor}, \eqref{GeodesicDistance} we have
\bal
  n^i(\sigma) n^i(\tau) &=
  1 - \frac{1}{2} \left( \partial_a n^i \partial_b n^i \right) \eta^a \eta^b +
  \ho[3]{\eta}\,,\\
  \sqrt{h(\tau)} &=
  1 - \frac{1}{6} \RicciTensor^\Sigma_{a b} \eta^a \eta^b + \ho[3]{\eta} \,,
  \\
  \norm{x(\sigma)-x(\tau)}^2 & = \eta^a \eta_a - \frac{1}{12}\IIFundForm_{a b}
\cdot \IIFundForm_{c d} \eta^a \eta^b \eta^c \eta^d + \ho[5]{\eta}\,.
\label{eqn:FTexpansion}
\eal

The integral computing the density of the scalar contribution to $\log \vev{V_\Sigma}$ is then
\beq
\frac{1}{4\pi^2} \int \frac{\diff^2 \eta }{|\eta|^4} 
  \left[ 1 - \left( \frac{1}{6} R^\Sigma_{ab} + \frac{1}{2} \partial_a n^i
    \partial_b n^i \right)\eta^a \eta^b 
  +\frac{1}{6|\eta|^2}\IIFundForm_{a b}\cdot \IIFundForm_{c d} \eta^a \eta^b \eta^c \eta^d
+ \ho[3]{\eta} \right].
\label{eqn:FTscalarcontrib}
\eeq
Using polar coordinates $\eta^a = \eta\, e^a(\varphi)$, where $e$ is a 2d 
unit vector, and the identities 
\beq
\int_0^{2\pi} \diff \varphi\, e^a e^b = \pi \delta^{ab}, \qquad
\int_0^{2\pi} \diff\varphi\, e^a e^b e^c e^d = \frac{\pi}{4} \left(
\delta^{ab}\delta^{cd} + \delta^{ac}\delta^{bd} + \delta^{ad} \delta^{bc}
\right),
\label{eqn:FTpolaridentities}
\eeq
we are left with the radial integral, for which we introduce the cutoff $\epsilon$
\bal
&\frac{1}{2\pi} \int_\epsilon \frac{\diff \eta}{\eta^3} \Big( 1 
-\frac{ \eta^2}{48} \left(4\RicciScalar^\Sigma + 12 \cAnomaly 
-H^2 - 2\IIFundForm^{ab} \cdot \IIFundForm_{ab}\right) + \ho[3]{\eta} \Big) \\
&= \frac{1}{4 \pi\epsilon^2} + \frac{1}{32 \pi} \left( 2\RicciScalar^\Sigma -H^2 
+4 \cAnomaly \right) \log \epsilon + \fin.
\label{eqn:scalarfinal}
\eal
To get the expression in the second line we also used the Gauss-Codazzi
equation~\eqref{eqn:GaussCodazzi}.

The calculation of the contribution of the 2-form field is very similar. Expanding the tensor
structure, we have
\beq
\frac{1}{2}\vev{B^+(\sigma) B^+(\tau)} = 
\frac{1}{8\pi^2}\frac{\delta_{\mu\rho}\delta_{\nu\sigma}}{|x(\sigma)-x(\tau)|^4}\,
\diff x^\mu(\sigma) \wedge \diff x^\nu(\sigma) \otimes
\diff x^\rho(\tau) \wedge \diff x^\sigma(\tau)\,.
\eeq
In terms of $\eta^a$, the differential forms read (see \eqref{EmbeddingNormalCoordinates})
\bal
\label{eqn:DifferentialFormStructure}
    \diff x^\mu \wedge \diff x^\nu \big\rvert_{\sigma} &= 
  \varepsilon^{ab} v_a^\mu v_b^\nu \diff^2 \eta\,, \\
  \diff x^\rho \wedge \diff x^\sigma \big\rvert_{\tau} &= \varepsilon^{cd} 
  \bigg( v_{\mathstrut c}^{[\rho} v_d^{\sigma]} + 2 v_{\mathstrut c}^{[\rho} v_{d e}^{\sigma]} \eta^e 
  + \left( v_{\mathstrut c e}^{[\rho} v_{d f}^{\sigma]} + v_{\mathstrut c}^{[\rho} v_{d e f}^{\sigma]} \right) \eta^e \eta^f 
  + \ho[3]{\eta} \bigg) \diff^2 \eta\,.
\eal
Collecting terms and introducing a radial cutoff as above, we find the contribution
\beq
\label{eqn:2formfinal}
-\frac{1}{4 \pi\epsilon^2} - \frac{1}{32 \pi} \left(
-2 \RicciScalar^\Sigma+ 3 H^2
 \right) \log \epsilon + \fin.
\eeq

Finally, combining \eqref{eqn:scalarfinal} and \eqref{eqn:2formfinal} we find that the quadratic 
divergences cancel and we are left with
\beq
  \log{\vev{V_\Sigma}} = \frac{1}{8 \pi} \log \epsilon \int_\Sigma \vol_\Sigma
  \left[ \RicciScalar^\Sigma
  - H^2 
  + \cAnomaly \right] +
  \fin.
  \label{eqn:FTanomaly}
\eeq
Comparing to \eqref{eqn:SCAnomaly}, we can read off the anomaly coefficients
\beq
a^{(1)}_1 = + \frac{1}{2}\,, \qquad
a_2^{(1)} = - \frac{1}{2}\,, \qquad
c^{(1)} = + \frac{1}{2}\,.
\label{eqn:FTanomalycoeff}
\eeq

As discussed above, since we do not know the contribution of the Weyl tensor to the $B$-field 
propagator, we cannot determine $b^{(1)}$. According to the conjecture
of~\cite{Bianchi:2019sxz} however, it should vanish. This relation is the
subject of work in progress~\cite{Drukker:2020atp}.

Equation \eqref{eqn:FTanomaly} differs from \eqref{eqn:SCAnomaly} by the absence of the $\tr P$ term, 
which also vanishes in flat space. Since $H^2$ doesn't 
vanish in flat space, it determines $a_2$ unambiguously and in curved space $H^2$ is 
necessarily accompanied by $4\tr P$, based on the general argument for the form of the anomaly 
reviewed in Section~\ref{sec:anomaly}.

Finally, we reiterate that, depending on the form of the quantisation condition, 
the result for the anomaly coefficients may be changed by a factor of $2$, see the discussion 
following~\eqref{eqn:Propagators}. In any case, the abelian theory should have 
surface operators with an integer multiple of $iB^+-n^i\Phi_i$ in \eqref{WilsonSurface_Definition}, 
and for all of them it is still true that $a^{(1)}_1 =-a_2^{(1)} =c^{(1)}$.

\subsection{Generalising the scalar coupling}

Note that the preceding calculation is applicable regardless of whether the
operator is locally BPS or not, so we may relax the condition $n^2 = 1$. In that
case the result for the anomaly coefficients is
\beq
a^{(1)}_1 = \frac{n^2 + 1}{4}\,.
\qquad
a^{(1)}_2 = - \frac{n^2 + 3}{8}\,,
\qquad
c^{(1)}= \frac{1}{2}\,.
\label{gen-n}
\eeq
If we replace $n^i \to i n^i$, we recover the surface operator studied
in~\cite{gustavsson:2004gj}. An operator with $n^2=0$ was also studied in~\cite{henningson:1999xi}, 
but assuming a non-self-dual 2-form. The anomaly coefficients computed 
in~\cite{henningson:1999xi,gustavsson:2004gj} are respectively twice and four
times the ones we obtain by substituting the values of $n$ in \eqref{gen-n}, due
to a difference in the overall normalisation of the propagator. 

It would be interesting to study this system in the large $n^2$ limit. This is similar to the 
``ladder'' limit of the cusped Wilson loop in ${\cal N}=4$ SYM in 4d first suggested in 
\cite{Correa:2012nk} which is related to a special scaling limit of that theory, dubbed the 
``fishnet'' model, which also has a 6d version~\cite{Gromov:2017cja}.

\section{Holographic description at large $N$}
\label{sec:hol}

The holographic calculation of the Weyl anomaly for surface operators was
pioneered by Graham and Witten in~\cite{graham:1999pm}. Here we present a
rewriting of their argument, which we also generalise slightly to include operators
extended on the $S^4$.

\subsection{Surface operators}

The $\mathcal{N} = (2,0)$ theory is described at large $N$ by 11d supergravity on
an asymptotically $AdS_7 \times S^4$ geometry~\cite{maldacena:1997re}
\beq
  \label{eqn:AdSmetric}
  \diff s^2 = \frac{L^2}{y^2} \left( \diff y^2 + g^{(0)} + g^{(1)} y^2 \right)
  + \frac{L^2}{4} g_{S^4}^{(0)} + \ho[2]{y}\,, \qquad
  L = \left( 8 \pi N \right)^{1/3} l_P\,,
\eeq
such that $g^{(0)}$ is the metric of the dual field theory%
\footnote{Or in the same conformal class.} 
and $g_{S^4}^{(0)}$ is the metric of $S^4$.

The background also includes $N$ units of $F_4$ flux
\beq
\label{S4flux}
  \frac{1}{\left( 2 \pi \right)^2 l_P^3} \int_{S^4} F_4 = 2 \pi N\,.
\eeq

The full form of the metric is determined by the supergravity equations of
motion in the presence of fluxes and by requiring the geometry to close smoothly in
the interior. While the latter requires nonlocal information, the near-boundary
expansion is fixed to the required order by local information about the boundary. 
Following~\cite{fefferman1985conformal, fefferman:2007rka}, the first term in this 
expansion was found in~\cite{graham:1999pm} as
\beq
  g^{(1)}_{\mu\nu} = -P_{\mu\nu}^{(0)}\equiv- P_{\mu\nu}\big|_{g = g^{(0)}}\,.
  \label{eqn:GWAdSmetriccontrib}
\eeq
At this order the $S^4$ is round, so to leading order the solution to
\eqref{S4flux} is simply
\beq
  F_4 = \frac{3}{8} L^3 \vol_{S^4}\,.
\eeq

The holographic description of the surface operators~\eqref{WilsonSurface_Definition} is by 
M2-branes anchored along $\Sigma$ on the boundary of $AdS$~\cite{maldacena:1998im}. Using 
$\hat{\Sigma}$
for the world-volume of the M2-brane, it has a boundary at $y=0$ with $\partial\hat{\Sigma}=\Sigma$.
The expectation value of the surface operators is then given by the minimum of the 
M2-brane action, reading (in Euclidean signature and with all fermionic terms suppressed)
~\cite{Bergshoeff:1987cm} 
\beq
  \label{eqn:GWaction}
\log{\vev{V_\Sigma}} \simeq 
- S_{\text{M2}} = 
-T_{\text{M2}}
  \int_{\hat{\Sigma}} \left( \vol_{\hat{\Sigma}} + \im A_3 \right),
  \qquad T_{\text{M2}} = \frac{1}{4\pi^2 l_P^3}
  =\frac{2N}{\pi L^3}\,,
\eeq
where  $T_{\text{M2}}$ is
the tension of the brane, proportional to $N$. $\vol_{\hat{\Sigma}}$ is the volume form calculated
from the induced metric and $A_3$ is the pullback of the 3-form potential.

\subsection{Local supersymmetry}

Before studying the M2-brane embeddings, let us note that the M2-brane
minimizing~\eqref{eqn:GWaction} is also locally supersymmetric. The supergravity
fields appearing there sit in the supergravity multiplet, which transform as
\bal
    \delta A_{MNP} &= -3 \bar{\varepsilon} \Gamma_{[ MN } \Psi_{P]}\,,\\
    \delta \Psi_M &= D_M \varepsilon + \frac{1}{288} \left(
    \tensor{\Gamma}{^{PQRS}_M} - 8 \Gamma^{QRS} \delta^{P}_M
    \right) F_{PQRS} \varepsilon\,,\\
    \delta E^{\bar{M}}_M &= \bar{\varepsilon} \Gamma^{\bar{M}} \Psi_M\,,
  \label{eqn:HoloSUSY}
\eal
where $E^{\bar{M}}_M$, $\Psi_M$ and $A_3$ are respectively the vielbein, gravitino and
3-form potential of $F_4$ ($\bar{M}=1,\dots,11$ is the frame index).
Using these transformations, the variation
of~\eqref{eqn:GWaction} is
\beq
  \delta_\varepsilon S = T_\text{M2} \int_{\hat{\Sigma}}
  \bar{\varepsilon} \left( \Gamma^{\hat{a}}
  - \frac{\im}{2} \varepsilon^{\hat{a} \hat{b} \hat{c}} 
  \Gamma_{\hat{b} \hat{c}} \right) \Psi_{\hat{a}}
  \vol_{\hat{\Sigma}} = 0\,.
\eeq
We here denote the coordinates on the world-volume by
$\hat{\sigma}^{\hat{a}}$.
The projector equation is then
\beq
  \bar{\varepsilon} \Pi_{-} = 0, \quad
  \Pi_- = \frac{1}{2} \left[ 1 - \frac{\im}{6} \varepsilon^{\hat{a} \hat{b} \hat{c}}
  \Gamma_{\hat{a} \hat{b} \hat{c}} \right].
\eeq
The projector is again half-rank, so that the M2-brane locally preserves half of
the supersymmetries (16 supercharges). These supercharges can be shown to agree
with the field theory BPS condition~\eqref{eqn:BPSXproj} on $\Sigma$ once we
decompose $x^M$ into coordinates on the boundary of $AdS$, $x^\mu$, and the
$S^4$ coordinates $n^i$.

\subsection{Holographic calculation}

To find the saddle points of the action~\eqref{eqn:GWaction}, we 
parametrise the M2-brane by $y, \sigma^a$ where $\sigma^a$ are coordinates for 
$\Sigma$. We then use the static gauge to 
describe the embedding by $\{u^{a'}(y,\sigma), n^i(y,\sigma)\}$, where 
$u^{a'}$ are the normal directions to the surface $\Sigma$ at $y=0$. In this 
setup, the boundary conditions are $u^{a'}(y=0,\sigma)=0$ and 
$n^i(y=0,\sigma)=n^i(\sigma)$ (where the right hand side has the $n^i$ from 
\eqref{WilsonSurface_Definition}).

Because the metric~\eqref{eqn:AdSmetric} 
diverges at the boundary of $AdS$, the volume element on the 
M2-brane diverges as $y^{-3}$, which leads to divergences in the action. 
Finding the shape of the embedding requires knowledge of the full surface 
and is generally a hard problem. But since we are only interested in the
logarithmically divergent part of the action, it is sufficient to solve the
equations of motion for small $y$. We do this perturbatively
following~\cite{graham:1999pm}, mirroring the solution of the background
supergravity equations above.

Using \eqref{eqn:GWAdSmetriccontrib}, the lowest order terms in the metric 
for our coordinates normal and tangent to the surface, are
\bal
    g_{ab}(y,\sigma,u) &= h_{ab} - P_{ab}^{(0)} y^2 +\partial_{a'}g^{(0)}_{ab}\Big|_{u = 0} u^{a'}
    + {\cal O}(y^4,u^2)\,,\\
    g_{a a'}(y, \sigma, u) &= {\cal O}(y^2,u)\,,\\
    g_{a'b'}(y,\sigma,u) &= g^{(0)}_{a'b'}\Big|_{u=0} + {\cal O}(y^2,u)\,.
\eal
Here $h_{ab} = g_{ab}^{(0)}\Big|_{u=0}$ is the metric on $\Sigma$. Note that away from $y=0$,
this metric depends on $u^{a'}$ (for $y \neq 0$, generically $u^{a'} \neq 0$),
as in the first line.

To write down the M2-brane action we need the induced metric
$\hat{h}_{ab} = \partial_a X^M\partial_b X^N g_{MN}$ (including also the $S^4$ directions). 
We expand the embedding coordinates as
\bal
    u^{a'}(y, \sigma) &= \mathcal{O}(y^2)\,,\\
    n^i(y, \sigma) &= n^i(\sigma) + \mathcal{O}(y^2)\,.
  \label{eqn:GWexpansion}
\eal
It is easy to check that higher order terms are not required. Then the $S^4$ metric 
can be replaced with $g_{S^4}^{(0)} = \delta_{ij} \diff n^i \diff n^j$ and the 
second fundamental form is $\IIFundForm^{a'}_{ab} = -\frac{1}{2}
g^{a'b'}\partial_{b'} g_{ab}$.

Dropping the explicit 
$\cO(y^\star)$ as well as the subscript $|_{u=0}$ along with the superscript ${}^{(0)}$,
since all the quantities are evaluated on the surface, we find
\bal
    \hat{h}_{y y} &\simeq \frac{L^2}{y^2}
    \left[ 1 + \partial_y u^{a'} \partial_y u^{b'} g_{a'b'} \right],\\
    \hat{h}_{a y} &\simeq 0\,,\\
    \hat{h}_{a b} &\simeq \frac{L^2}{y^2}
    \left[ h_{ab} + \left(-P_{ab}
      + \frac{1}{4} \partial_a n^i  \partial_b n^j \delta_{ij}\right) y^2
      -2 \IIFundForm^{a'}_{ab} u^{b'}g_{a'b'} \right].
\eal

The determinant of the metric is then
\beq
  \det{\hat{h}} \simeq \frac{L^6}{y^6}
  \left( 1 + \partial_y u^{a'} \partial_y u^{b'}g_{a'b'}  - 2 H^{a'}u^{b'} g_{a'b'} 
  + \left( - \tr P + \frac{1}{4} \cAnomaly \right) y^2 \right) \det{h}\,,
\eeq
while the pullback of the 3-form
\beq
  A_3 = \frac{1}{3!} A_{ijk}\, \diff n^i \wedge \diff n^j \wedge \diff n^k \sim
  \ho{y}\,,
\eeq
does not contribute to the divergences. We thus find the action
\begin{equation}
  S_{\text{M2}} \simeq \frac{L^3}{\left( 2 \pi \right)^2 l_P^3}
  \int_{\Sigma} \vol_\Sigma \int\limits_{y \ge \epsilon} \frac{\diff y}{y^3}
  \left[ 1 + \frac{1}{2} \left(\partial_y u^{a'}\right)^2 - H \cdot u  + \left(- 4\tr{P} + \cAnomaly \right)
    \frac{y^2}{8}  \right].
    \label{eqn:GWactiony}
\end{equation}
At order $\ho[2]{y}$, we need only solve for $u^{a'}(y)$, which has the
equation of motion
\beq
  y^3 \partial_y \left( y^{-3} \partial_y u^{a'} \right) + H_{a'} \simeq 0
  \qquad\Rightarrow\qquad
  u^{a'} \simeq \frac{1}{4} H^{a'} y^2\,.
\eeq
The action evaluated at the classical solution is then
\beq
  S_{\text{M2}} \simeq \frac{L^3}{\left( 2 \pi \right)^2 l_P^3}
  \int_{\Sigma} \vol_\Sigma \int\limits_{y \ge \epsilon} \frac{\diff y}{y^3}
  \left[ 1 - \frac{y^2}{8} \left(H^2 + 4 \tr{P} \right)
    + \frac{y^2}{8} \cAnomaly \right]
  \label{eqn:GWactionEff}
\eeq
where we see that the anomaly indeed takes the form~\eqref{eqn:SCAnomaly}. The
result is
\beq
  \log{\vev{V_\Sigma}} = \frac{N}{4 \pi} \log{\epsilon}
  \int_{\Sigma} \vol_\Sigma 
  \left[ - \left(H^2 + 4 \tr{P} \right) + \cAnomaly \right]
  \log{\epsilon} + \fin,
  \label{eqn:GWvev}
\eeq
where we discarded an irrelevant term proportional to
$\epsilon^{-2}$ (see the discussion below).

This result agrees with the original calculation of~\cite{graham:1999pm} and adds to it 
the coupling to $(\partial n)^2$. It is also consistent with the explicit calculation
of the $1/2$-BPS sphere~\cite{berenstein:1998ij}, for which the anomaly is $-4N$.
The anomaly coefficients at leading order in $N$ are then
\bal
a^{(N)}_1 &= \ho[0]{N} \,,&
  \qquad
  b^{(N)} &= \ho[0]{N} \,, \\
  a_2^{(N)} &= - N + \ho[0]{N} \,,&
\qquad
c^{(N)} &= + N + \ho[0]{N}\,.
\label{eqn:GWanomalycoeff}
\eal
As in the case of Wilson loops in ${\cal N}=4$ SYM in 4d, we expect this
holographic description to be correct in the locally BPS case when the scalar
couplings satisfy $n^2=1$.  Following \cite{Alday:2007he,
Polchinski:2011im}, the case of $n^2=0$ should
be described by the same surface inside $AdS_7$, but completely smeared over the
$S^4$. In this case we find the same result for the geometric anomaly coefficients 
as above, and, since the corresponding anomaly term vanishes, $c^{(N)}$ does not 
apply.

\subsubsection{Power-law divergence}
\label{sec:powerlaw}

Note that in addition to the log divergence in~\eqref{eqn:GWvev}, 
\eqref{eqn:GWactionEff} produces also a power-law divergence
\beq
  \frac{L^3}{\left( 2 \pi \right)^2 l_P^3} \frac{\text{Area}(\Sigma)}{2 \epsilon^2}\,.
  \label{eqn:GWpowerlaw}
\eeq
While such divergences can be removed by the addition of a local counter-terms, 
in the field theory result \eqref{eqn:FTanomaly}, they cancelled without extra 
counter-terms (for the locally BPS operator). 

A more elegant way of eliminating the power law divergences also in this holographic 
calculation follows the example of the locally BPS Wilson loops~\cite{drukker:1999zq}. 
A careful treatment of the boundary conditions suggests that the natural action is a
Legendre transform of~\eqref{eqn:GWaction}, which differs from the action we
used by a total derivative. This modification does not change the equations of
motion, but gives a contribution on the boundary, where it precisely cancels the
divergence above.

By looking at the M5-brane metric before the decoupling limit, we can identify 
the coordinate to use in the transform as $r^i=L^3 n^i/2 y^2$. Defining 
its conjugate momentum by differentiating with respect to the boundary 
value of the coordinate (where $y=\epsilon$)
\beq
  p_i(\sigma) = \frac{\delta S[x^\mu,r^i]}{\delta r^i}
  =-\frac{\epsilon^3n^i}{L^3}\frac{\delta S[x^\mu,n^i,\epsilon]}{\delta \epsilon}
  =\frac{\epsilon^3n^i}{L^3}\frac{L^3}{(2\pi)^2l_P^3}\left(\frac{1}{\epsilon^3}+\cO\left(\frac{1}{\epsilon}\right)
  \right).
\eeq
In the last equality we used the value of the classical action~\eqref{eqn:GWactionEff}, undoing 
the integration, so the classical Lagrangian density.

The Legendre transformed action is then
\beq
\tilde{S}\left[ x^\mu, p^i \right] 
= S\left[ x^\mu, r^i \right] -\int_{\Sigma} p_i r^i \vol_\Sigma
= S\left[ x^\mu, n^i,\epsilon \right]
-\frac{L^3}{2(2\pi)^2l_P^3\epsilon^2}\int_{\Sigma}  \vol_\Sigma\,.
\eeq
The last term exactly cancels the power law divergence in~\eqref{eqn:GWpowerlaw}.

\section{Surfaces with singularities}
\label{sec:singularity}

An interesting class of surface operators that has received some attention
recently is surfaces with conical singularities. For these surfaces, it was
found that the regularised expectation value typically diverges
as~\cite{klebanov:2012yf,myers:2012vs,bueno:2015lza,dorn:2016bkd}
\beq
  \log{\vev{V_{\Sigma_c}}} \sim A \log^2 \epsilon + \mathcal{O}(\log{\epsilon})\,.
\eeq

Let us consider a conical defect (on flat space) of the form
\beq
  x^\mu(r, s) = r \gamma^\mu(s)\,, \qquad 
  \gamma^2 = 1\,,\qquad
n^i(r,s)=\nu^i(s)\,.
  \label{eqn:conical}
\eeq
We allow here also a ``conical singularity'' in the scalar couplings, which has $s$ dependence 
even as $r\to0$. It is possible to also allow $x^\mu$ and $n^i$ to have higher order terms in $r$, 
but since those lead to subleading divergences, they are unimportant.

We can try to use the usual formula for the anomaly \eqref{eqn:SCAnomaly} by plugging in 
the geometric invariants
\beq
\label{conecurv}
  \RicciScalar^{\Sigma} = \Omega\delta(r)\,, \qquad
  H^2 = \frac{\kappa^2 - 1}{r^2}\,, \qquad
  \left( \partial n \right)^2 =
  \frac{( \partial_s \nu)^2}{r^2}\,,
\eeq
where $\Omega$ is the deficit angle, $\kappa = {\ddot \gamma}^2/|\dot\gamma|^2$ 
is the curvature of $\gamma$. Plugging into~\eqref{eqn:SCAnomaly}, the Ricci scalar 
gives a finite contribution, but $H^2$ and
$(\partial n)^2$ diverge as $r \to 0$. Introducing a cutoff $\hat\epsilon$ on the $r$ integration, 
this gives
\beq
\label{eq:naive}
\frac{1}{4 \pi} \log\epsilon\log\hat \epsilon \int_\gamma
  a_2 \left(1 - \kappa^2(s) \right) - c (\partial_s \nu)^2
  \diff s + \ho{\log{\epsilon}}\,.
\eeq

This expression is a bit naive, as we should treat all divergences on the same footing and 
should identify $\hat\epsilon=\epsilon$. But then we should not use~\eqref{eqn:SCAnomaly}, 
rather go back one step and regularise the divergences that gave rise to the original 
$\log\epsilon$ divergence while also applying it to the $r$ integration. As we show below, 
this leads to the expression in \eqref{eq:naive} with 
$\log\epsilon\log\hat\epsilon\to\frac{1}{2}\log^2\epsilon$. In both the free field case and 
the holographic realisation this factor of $1/2$ is a simple consequence of the usual coefficient 
of the quadratic term in the Taylor expansion, or in other words of an integral of the form 
$\int \log r\,\diff\log r$. 

This factor of $1/2$ was noticed already in the calculations 
of~\cite{klebanov:2012yf,myers:2012vs} 
and justified in~\cite{dorn:2016bkd} by a careful treatment of the holographic 
calculation, which is repeated below. It was also studied in the context of
defect CFT in~\cite{bianchi:2015liz}.
We think that the comparison of this to the 
free-field calculation and the universal nature of our result further elucidates this 
mismatch from the naive expectation. Our calculation is also more generic, for 
allowing arbitrary conical singularities and incorporating the scalar singularities too.

Beside this factor $1/2$, it is interesting to compare the $\log^2{\epsilon}$
divergence of surface operators to the $\log{\epsilon}$ divergence of cusped Wilson
loops.  In $\cN = 4$ SYM, the cusp anomalous
dimension is a complicated function of the opening angle $\phi$. At small
angles, it is related to the Bremsstrahlung function, which encodes the
radiation emitted by heavy probe particles.  It is therefore an
interesting quantity to compute, and the exact Bremsstrahlung function has been
obtained using supersymmetric localization in~\cite{correa:2012at}.

In constrast, the expression~\eqref{eq:gaugecon} is not an approximation for small
angles, but the exact result. The relation to physical quantities is unclear
as well. It would be interesting to interpret it as a Bremsstrahlung
function, but computing the radiation emitted by a probe string in 6d would
require a more careful treatment of the self-dual field strength.

We should also note, as already noticed in \cite{myers:2012vs }, that surfaces with 
``creases'', i.e. co-dimension one singularities, do not lead to additional
$\log^2{\epsilon}$ divergences and the expression~\eqref{eqn:SCAnomaly} can be
immediately applied to them.

\subsection{Field theory}

Here we do not rely on~\eqref{eqn:FTanomaly}, but go further back to 
where the $\log{\epsilon}$ arises from an integral of the form~\eqref{eqn:scalarfinal}
\beq
\int_\epsilon^\rho \frac{\diff \eta}{\eta} = - \log \epsilon + \fin,
\eeq
where $\eta$ is a radial coordinate around the point $x$, and $\rho$ is an IR cutoff 
related to the overall size of the surface, or at least a large smooth patch where 
we defined our local coordinate. Near the cone the smooth patch is bounded by the 
distance from $x$ to the apex, which we denote by $r$. The integral instead gives
\beq
  \int_\epsilon^r \frac{\diff \eta}{\eta} = -\log{\frac{\epsilon}{r}}\,.
\eeq
With this careful treatment of the log, we can go back to~\eqref{eqn:SCAnomaly}, 
plug in the expressions from~\eqref{conecurv} and integrate over $r$ and 
with the same UV cutoff to find
\bal
\label{eq:gaugecon}
  \log{\vev{V_\Sigma}} &= - \frac{1}{4 \pi}
  \int_\gamma \diff s
  \int\limits_{\epsilon} \frac{\diff r}{r}
  \left[ a_2 \left( 1 - \kappa^2 \right) - c(\partial_s n)^2 \right]
  \log{\frac{\epsilon}{r}} + \fin
\\&
=\frac{1}{8 \pi} \log^2\epsilon\int_\gamma
\left[ a_2 (1 - \kappa^2(s)) - c (\partial_s \nu)^2 \right]
  \diff s + \ho{\log{\epsilon}}\,.
\eal

\subsection{Holography}

The derivation in holography is similar. We first note that conformal symmetry fixes the form 
of the solution as
\beq
  y(r,s) = r u(s) 
\eeq
To get to~\eqref{eqn:GWvev}, we integrate over $y$, but the conformal ansatz suggests to impose 
the range $\epsilon\leq y\leq ru_\text{max}$. Plugging the curvatures from \eqref{conecurv} into 
equation~\eqref{eqn:GWvev} we arrive at
\beq
\label{eqn:cone_holography}
  \log{\vev{V_\Sigma}} = - \frac{1}{4 \pi}
  \int_\gamma \diff s
  \int\limits_{\epsilon} \frac{\diff r}{r}
  \left[ a_2 (1 - \kappa^2) - c (\partial_s n)^2 \right]
  \log{\frac{\epsilon}{r u_\text{max}(s)}} + \fin.
\eeq
which again gives the $\log^2\epsilon$ divergence with the same $1/2$ prefactor, as in the 
field theory~\eqref{eq:gaugecon}.

\subsection{Example: circular cone}

\begin{figure}[h]
  \centering
  \includegraphics[scale=1]{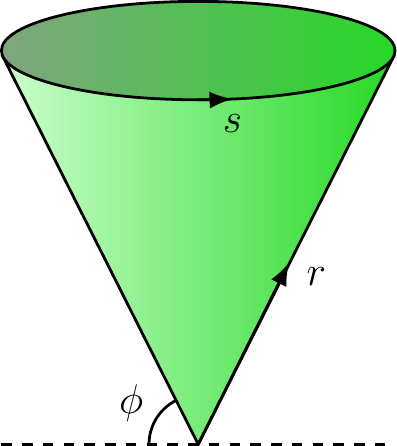}
  \hspace{0.2\textwidth}
  \includegraphics[scale=1]{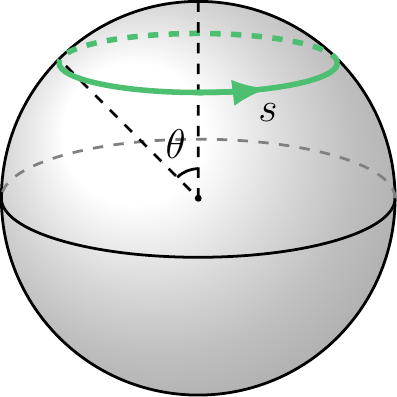}
  \caption{On the left, the surface wraps a (circular) cone with a deficit
  angle $\phi$. On the right, the scalar coupling follows a circle at
  angle $\theta$ on $S^2$. For a fixed $r$, we have a curve that simultaneously 
  traces the circles $\gamma(s)$ and $n^i(s)$.}
  \label{fig:cone}
\end{figure}

As a simple example of a singular surface we compute explicitly the anomaly
of a cone. Denoting the deficit angle by $\phi$ (see figure~\ref{fig:cone}) 
and including an internal angle $\theta$ for the scalar coupling $n^i$, 
we parametrise the cone as follows
\beq
\label{ConeEmbedding}
\gamma^\mu(s) = 
\begin{pmatrix}
 \cos\phi \sin s \\
 \cos\phi \cos s \\
 \sin\phi
\end{pmatrix}, \qquad
n^i(s) = 
\begin{pmatrix}
 \sin\theta \sin s \\
 \sin\theta \cos s \\
 \cos\theta
\end{pmatrix},
\qquad
0\leq r\,,\quad 0\leq s<2\pi\,.
\eeq
The conformal invariants are explicitly
\beq
  \kappa^2 = \frac{1}{\cos^{2}\phi}\,, 
  \qquad 
  \left( \partial_s n \right)^2 =
  \frac{\sin^2{\theta}}{\cos^2{\phi}}\,.
\eeq
The divergence is then
\beq
  \log{\vev{V_\Sigma}} = - \frac{a_2 \sin^2{\phi} + c \sin^2{\theta}}{4
    \cos{\phi}} \log^2{\epsilon} + \ho{\log{\epsilon}}\,.
\eeq
Notice that as long as the anomaly coefficients satisfy the relation $a_2 = - c$,
which we have shown to hold in the abelian and large $N$ case, the anomaly 
vanishes for configurations $\theta = \pm \phi$, which correspond generically 
to $1/8$-BPS configurations.

\section{Conclusion}
\label{sec:conclusion}

In this paper we calculated the anomaly coefficients of locally supersymmetric surface 
operators in the $\cN=(2,0)$ theory in 6d, refining and generalising the calculations 
of~\cite{graham:1999pm, gustavsson:2004gj}. 
We first introduced a new anomaly coefficient $c$ \eqref{eqn:SCAnomaly} 
arising from non-constant dependence on the internal R-symmetry directions. 
These are explicit scalar couplings in the abelian theory and motion on 
$S^4$ in the holographic realisation.

We then presented an explicit calculation for the abelian theory and for the large $N$ limit 
(using holography). The results are in equations \eqref{eqn:FTanomalycoeff} and 
\eqref{eqn:GWanomalycoeff}. Although we are not able to compute
the anomaly coefficient $b$ at $N=1$ because we do not know the general curved
space propagator for the self-dual 2-form, we found the others in both cases.

Making all $N$ conjectures based on the asymptotics is a fool's errand, which we
carefully tread. This is especially true given that the abelian theory is not the same
as the $A_{N-1}$ theory at $N=1$, since the latter is the empty theory.
Nevertheless, in both cases we see that $a_2=-c$, and we expect this to 
hold generally. The argument
is based on the BPS Wilson loops of \cite{zarembo:2002an}, where $n^i$ is
parallel to $\dot x^\mu$ and which have trivial expectation values. If we uplift
them to the 6d theory we expect to find surface operators with no anomaly (and
no finite part as well). These operators satisfy $H^2=(\partial n)^2$ and indeed
they do not contribute to the anomaly\footnote{%
In the uplift we find only
surfaces with trivial topology, so the anomaly vanishes regardless of $a_1$.} 
for $a_2=-c$. 
A proof of this relation as well as properties of $b$, based on defect CFT
techniques, will be presented elsewhere \cite{Drukker:2020atp}.

Two more results are the formalism for regularising surface operators presented 
in Appendix~\ref{sec:GeodesicPointSplitting} and the expression for the divergences 
due to conical singularities over arbitrary curves in Section~\ref{sec:singularity}.

All our calculations are for a surface operator in the fundamental representation. 
It is expected that $1/2$-BPS surface 
operators are classified by representations of the $A_{N-1}$ algebra of the theory. 
At large $N$ this is proven, since the   
asymptotically $AdS_7 \times S^4$ solutions of 11d supergravity 
preserving the symmetry algebra of $1/2$-BPS surface operators can be classified  in terms of 
Young diagrams~\cite{DHoker:2008rje, bachas:2013vza}.

A calculation of anomalies of surface operators in arbitrary representation, based on the bubbling 
geometries and holographic entanglement
entropy was undertaken in~\cite{Estes:2018tnu}. If we assume $b = 0$, then for a
the fundamental representation, their result reads
\beq
  a_1^{(N)} = \frac{1}{2}-\frac{1}{2N}\,,\qquad
  a_2^{(N)} = -N + \frac{1}{2} +\frac{1}{2N}\,.
\eeq
This is supported by an independent calculation using the superconformal
index~\cite{Chalabi:2020iie}. In the large $N$ limit, our
result~\cite{graham:1999pm} indeed agrees with theirs. 
These calculations do not determine the remaining anomaly coefficients in generic
representations. But if we believe the $b = 0$ conjecture of~\cite{Bianchi:2019sxz}
and our argumentation above for $c = -a_2$, this fixes the remaining ones.

It would be interesting to reproduce these finite $N$ corrections using other 
methods as well as do direct holographic calculations for higher-dimensional 
surface operators.

The anomalies studied here are the most basic properties of surface operators, but 
finding them is only a first step in understanding these observables and the mysterious 
theory they belong to. Planar/spherical surface operators preserve part of the conformal 
group (and with the scalar coupling also half the supersymmetries) and their deformations 
behave like operators in a defect CFT. A natural next step is to study the defect CFT 
data: spectrum and structure constants.

Another natural question is the classification of globally BPS surface operators 
(and local operators within the surface operators associated with their
deformations) beyond the case of the plane/sphere.

We hope to report progress on these questions in the near future. 
\subsection*{Acknowledgments}

We would like to thank 
Lorenzo Bianchi,
Robin Graham,
Nikolay Gromov,
Chris Herzog,
Elias Kiritsis,
Neil Lambert,
Ki-Myeong Lee,
Andy O'Bannon, 
Ronnie Rodgers and
Itamar Shamir
for interesting discussions.
ND would like to thank the University of Modena and Reggio Emilia, the Pollica Summer Workshop, 
the Hebrew University, CERN and EPFL Lausanne for hospitality in the course of this work. His work is 
supported by an STFC grant number  ST/P000258/1. The Pollica Summer workshop
was supported in part by the Simons Foundation (Simons Collaboration on the Non-perturbative Bootstrap) 
and in part by the INFN. The author is grateful for this support.
MT acknowledges the support of the Natural Sciences and Engineering Research Council 
of Canada (NSERC).  
Cette recherche a \'et\'e financ\'ee par le Conseil de recherches en sciences naturelles 
et en g\'enie du Canada (CRSNG).

\appendix
\section{Conventions and notation}
\label{sec:convention}

In this paper we work in Minkowski space with mostly positive signature. We make
use of the following indices:
$$
\label{indices}
  \begin{tabular}{l | l}
    Index & Usage \\ \hline
    $M = 1, \dots, 11$ & 11d spacetime vector $X^M$\\
    $A = 1, \dots, 32$ & 11d spinors\\
    $\mu = 1, \dots, 6$ & 6d spacetime vectors $x^\mu$\\
    $\alpha\ (\dot{\alpha}) = 1, \dots, 4$ & 6d chiral (antichiral) spinors\\
    $i = 1, \dots, 5$ & R-symmetry vectors\\
    $\check{\alpha} = 1,\dots,4$ & R-symmetry spinors\\
    $a' = 1,\dots,4$ & spacetime vectors orthogonal to the surface\\
    $a = 1,2$ & worldsheet coordinates $\sigma^a$\\
    $\hat{a} = 1,2,3$ & worldvolume coordinates $\hat{\sigma}^{\hat{a}}$\\
  \end{tabular}
$$

Our usage of spinors is restricted to the supersymmetry
transformations~\eqref{eqn:susy} and~\eqref{eqn:HoloSUSY} but we include our
conventions for completeness. In general we follow the NW-SE convention for
indices summation
\beq
  \bar{\Phi} \Psi \equiv \bar{\Phi}^A \Psi_A\,,
\eeq
The conjugate and transpose act as
\beq
  \left( \Psi_A \right)^* = \left( \Psi^* \right)^A\,, 
  \qquad 
  \left( \mathcal{C}^{AB} \right)^T = \mathcal{C}^{BA}\,.
\eeq
Below we detail the properties of gamma matrices in $d=11$ and $d=6$, and we
state the reality condition on spinors. More details can be found
in~\cite{claus:1997cq} and references therein.

\subsection{$d=11$ Clifford algebra}

The 11d Clifford algebra is generated by the set of matrices
$\tensor{\left(\Gamma_M\right)}{_A^B}$ satisfying
\begin{align}
  \left\{ \Gamma_M, \Gamma_N \right\} = 2 \eta_{MN}.
\end{align}
Here for readability $M$ is used for flat spacetime, unlike~\eqref{eqn:HoloSUSY}
where it denotes curved spacetime.

The matrices may be chosen such that $\Gamma_0^\dagger = - \Gamma_0$ is
antihermitian while the others are hermitian
$\Gamma_M^\dagger = \Gamma_M\ (M \neq 0)$.
In addition, there is an orthogonal, real anti-symmetric matrix
$\mathcal{C}_{AB}$ such that
$\Gamma_M \mathcal{C} = - \left( \Gamma_M \mathcal{C} \right)^T$.
$\mathcal{C}$ naturally defines a real structure by relating $\Psi$ and
$\Psi^\dagger$ as
\begin{align}
  \bar{\Psi} \equiv - i \Gamma_0 \Psi^\dagger = \mathcal{C}^\dagger \Psi.
\end{align}
This is the Majorana condition.

\subsection{$d=6$ Clifford algebra}

An easy way to construct the 6d Clifford algebra is to decompose $\Gamma_M =
\left\{ \Gamma_\mu, \Gamma_i \right\}$ by introducing a chirality matrix
$\Gamma_* = \Gamma_0 \Gamma_1 \Gamma_2 \Gamma_3 \Gamma_4 \Gamma_5$.
The matrices are then (in the chiral basis)
\beq
  \Gamma_\mu = \begin{pmatrix}
    0 & \bar{\gamma}_\mu\\
    \gamma_\mu & 0\\
  \end{pmatrix} \otimes I_{4}\,, 
 \qquad
  \Gamma_i = \begin{pmatrix}
    -I_{4} & 0\\
    0 & I_{4}\\
  \end{pmatrix} \otimes \check{\gamma}_i\,, 
  \qquad
   \Gamma_* = \begin{pmatrix}
    -I_{4} & 0\\
    0 & I_{4}\\
  \end{pmatrix} \otimes I_4\,,
\eeq
where the algebra is
\beq
  \bar{\gamma}_\mu \gamma_\nu + \bar{\gamma}_\nu \gamma_\mu = 2 \eta_{\mu\nu}\,, 
  \qquad
   \gamma_\mu \bar{\gamma}_\nu + \gamma_\nu \bar{\gamma}_\mu = 2 \eta_{\mu\nu}\,, 
   \qquad
    \left\{ \check{\gamma}_i, \check{\gamma}_j \right\} = 2 \delta_{i j}\,.
\eeq
Since $\gamma_\mu$ and $\check{\gamma}_i$ commute, they define independent
spinor representations. Explicitly, we decompose $A = (\dot{\alpha}
\oplus \alpha) \otimes \check{\alpha}$, so that
the indices are $\tensor{\left( \gamma_\mu
\right)}{_\alpha^{\dot{\beta}}}$, $\tensor{\left( \bar{\gamma}_\mu
\right)}{_{\dot{\alpha}}^\beta}$ and
$\tensor{(\check{\gamma}_i)}{_{\check{\alpha}}^{\check{\beta}}}$. The chiral
and antichiral representations are related through
\beq
  \bar{\gamma}_\mu^\dagger = \gamma_0 \bar{\gamma}_\mu \gamma_0
   \Rightarrow \left\{ \begin{array}{ll}
      \bar{\gamma}_0^\dagger = -\gamma_0\,, &\\
      \bar{\gamma}_\mu^\dagger = \gamma_\mu\,, \qquad& \mu \neq 0\,.
    \end{array} \right.
\eeq
The chirality operator gives 2 additional constraints
\beq
  \gamma_{012345}
  = I\,, 
  \qquad
  \bar{\gamma}_{012345}
  = -I\,,
\eeq
with $\gamma_{\mu\nu \dots \rho} \equiv \gamma_{[\mu} \bar{\gamma}_{\nu} \dots
\gamma_{\rho]}$ the antisymmetrised product of $\gamma$-matrices.%
\footnote{(Anti-)symmetrisation is understood with the appropriate combinatorial factors, 
i.e. $A_{[a b]} = \frac{1}{2} A_{ab} - A_{ba} $.} 
The charge conjugation matrix takes the form
\beq
  \mathcal{C}_{AB} = \begin{pmatrix}
    0 & c_{\dot{\alpha} \beta}\\
    c_{\alpha \dot{\beta}} & 0\\
  \end{pmatrix} \otimes \Omega_{\check{\alpha} \check{\beta}}\,, 
  \qquad
  c \equiv c_{\dot{\alpha} \beta}\,,
\eeq
and is used to lower (or raise) spinor indices. The matrix $\Omega_{\check{\alpha}
\check{\beta}}$ is the real, antisymmetric symplectic metric of
$\mathfrak{sp}(4)$ and $c$ is unitary:
\beq
  c^\dagger c = c^{\alpha \dot{\alpha}} c_{\dot{\alpha} \beta}
  = \delta^{\beta}_{\alpha}\,, 
  \qquad
  c^* c^T = c^{\dot{\alpha} \alpha} c_{\alpha \dot{\beta}}
  = \delta^{\dot{\beta}}_{\dot{\alpha}}\,, 
  \qquad
  \Omega^\dagger \Omega =
  \Omega^{\check{\alpha} \check{\beta}} \Omega_{\check{\beta} \check{\gamma}}
  = \delta^{\check{\alpha}}_{\check{\gamma}}\,.
\eeq
They satisfy
\beq
  (\gamma_\mu c) = - \left(\gamma_\mu c \right)^T\,, \qquad
  \left( \bar{\gamma}_\mu c^T \right)
  = - \left( \bar{\gamma}_\mu c^T \right)^T\,, \qquad
  (\check{\gamma}_i \Omega) = - \left(\check{\gamma}_i \Omega \right)^T\,.
\eeq
A representation of this algebra is given by
\begin{align}
  \gamma_0 &= \bar{\gamma}_0 = i I_2 \otimes I_2\,,&\qquad
  \gamma_1 &= -\bar{\gamma}_1 = -i \sigma_1 \otimes I_2\,,&\qquad
  \gamma_2 &= -\bar{\gamma}_2 = -i \sigma_2 \otimes I_2\,,
\nonumber\\
  \gamma_3 &= -\bar{\gamma}_3 = i \sigma_3 \otimes \sigma_1\,,&\qquad
  \gamma_4 &= -\bar{\gamma}_4 = i \sigma_3 \otimes \sigma_2\,,&\quad
  \gamma_5 &= -\bar{\gamma}_5 = -i \sigma_3 \otimes \sigma_3\,,
\nonumber\\
  \check{\gamma}_1 &= \sigma_1 \otimes \sigma_2\,,\quad
  \check{\gamma}_2 = \sigma_2 \otimes \sigma_2\,,\hskip-6mm&
  \check{\gamma}_3 &= \sigma_3 \otimes \sigma_2\,,\quad
  \check{\gamma}_4 = I_2 \otimes \sigma_1\,,\hskip-6mm&
  \check{\gamma}_5 &= I_2 \otimes \sigma_3,
\nonumber\\
  c &= -c^T = \sigma_1 \otimes i \sigma_2\,,&\qquad
  \Omega &= i \sigma_2 \otimes I_2\,.\hskip-3cm
  \label{eqn:gammarep}
\end{align}

\subsection{Symplectic Majorana condition}

In 6d the spinor $\Psi$ decomposes into a chiral and an antichiral 6d spinor as
\beq
  \Psi_A = \begin{pmatrix}
    \bar{\chi}_{\dot{\alpha} \check{\alpha}}\\
    \psi_{\alpha \check{\alpha}}\\
  \end{pmatrix}, 
  \qquad
  \bar{\Psi}^A \equiv \begin{pmatrix}
    -i (\psi^\dagger)^{\alpha \check{\alpha}}
    \tensor{(\gamma_0)}{_\alpha^{\dot{\alpha}}} &
    \quad
    -i (\bar{\chi}^\dagger)^{\dot{\alpha} \check{\alpha}}
    \tensor{(\bar{\gamma}_0)}{_{\dot{\alpha}}^\alpha}
  \end{pmatrix} \equiv \begin{pmatrix}
    \bar{\psi}^{\dot{\alpha} \check{\alpha}} &\ 
    \chi^{\alpha \check{\alpha}}
  \end{pmatrix}.
\eeq
The Majorana condition on $\Psi$ then translates to
\beq
  \label{eqn:symplecticMajorana}
  \chi^{\alpha \check{\alpha}} =
  (c^\dagger \Omega^\dagger \bar{\chi})^{\alpha \check{\alpha}} =
  (c \Omega \bar{\chi})^{\alpha \check{\alpha}} \,, \qquad
  \qquad
  \bar{\psi}^{\dot{\alpha} \check{\alpha}} =
  (c^* \Omega^\dagger \psi)^{\dot{\alpha} \check{\alpha}} =
  -(c \Omega \psi)^{\dot{\alpha} \check{\alpha}}\,,
\eeq
where in the second equality we use the properties of our representation.
The inclusion of the symplectic form $\Omega$ in \eqref{eqn:symplecticMajorana} 
is the reason these equations are known as the \emph{symplectic Majorana condition}. 
The spinors $\bar{\varepsilon}^0$, $\varepsilon^1$, and $\psi$ in \eqref{eqn:susy} are 
of this type.

\section{Geometry of submanifolds}
\label{sec:GeometrySubmanifold}

In this appendix we assemble the geometry results used throughout the main text
and in Appendix~\ref{sec:GeodesicPointSplitting}.
Sections~\ref{sec:RiemannCurvature} and~\ref{sec:ExtrinsicCurvature} contain our
conventions for Riemann curvature and the definition of the second fundamental
form of an embedded submanifold as well as some standard results relating the
two. In Section~\ref{sec:NormalCoordinates} the second fundamental form is
related to the coefficients of the normal coordinate expansion of the embedding.

\subsection{Riemann curvature}
\label{sec:RiemannCurvature}

We adopt the convention where the Riemann tensor is defined as
\beq
\RiemannTensor\indices{^\mu_\nu_\rho_\sigma} = \partial_\rho \Gamma^\mu_{\nu \sigma} 
- \partial_\sigma \Gamma^\mu_{\nu \rho} + \Gamma^\mu_{\rho \lambda} \Gamma^\lambda_{\nu \sigma}
- \Gamma^\mu_{\sigma \lambda} \Gamma^\lambda_{\nu \rho}\,.
\eeq
It is convenient to split it into a conformally invariant Weyl
tensor $W_{\mu\nu\rho\sigma}$ and the Schouten tensor $P_{\mu\nu}$,
\begin{gather}
  \label{eqn:Schouten}
  P_{\mu \nu} = \frac{1}{d-2} \left( \RicciTensor_{\mu \nu} -
  \frac{\RicciScalar}{2(d-1)} g_{\mu \nu} \right),\\
  W_{\mu \nu \rho \sigma} = \RiemannTensor_{\mu \nu \rho \sigma} - g_{\mu \rho} P_{\nu \sigma}
  + g_{\mu \sigma} P_{\nu \rho} + g_{\nu \rho} P_{\mu \sigma} - g_{\nu \sigma} P_{\mu \rho}\,.
\end{gather}

\subsection{Extrinsic curvature}
\label{sec:ExtrinsicCurvature}

We define the second fundamental form to be
\beq
  \IIFundForm_{a b}^{\mu} = \left(\partial_a \partial_b x^\lambda 
  + \partial_a x^\rho \partial_b x^\sigma \Gamma^\lambda_{\rho\sigma}\right) 
  \left(\delta_\lambda^\mu - g_{\kappa\lambda}\partial^c x^\kappa \partial_c
  x^\mu \right).
  \label{eqn:IIFundForm}
\eeq
The second part is the projector to the components orthogonal to the surface
(defined by its embedding $x^\mu(\sigma)$), while the first part is the action
of the covariant derivative on the (pullback) of $x^\lambda(\sigma)$.  The mean
curvature vector is then
\beq
H^\mu = h^{ab} \IIFundForm^\mu_{ab}\,.
\eeq
These invariants are related to the intrinsic curvature of $\Sigma$ and $M$ by the Gauss-Codazzi 
equation
\beq
\label{Gauss-Codazzi}
\RiemannTensor^\Sigma_{abcd} = \RiemannTensor^M_{abcd} 
+ 2 \IIFundForm^\mu_{a[b}  \IIFundForm^\nu_{c]d} g_{\mu \nu} \,.
\eeq
Contracting twice with $h^{-1}$ and expanding the Riemann tensor in terms of
the Weyl and Schouten tensors, we obtain
\beq
  \left( H^2 + 4 \tr{P} \right) =
  2 \RicciScalar^{\Sigma} + 2 \tr{\tilde{\IIFundForm}^2} - 2 \tr{W}\,,
  \label{eqn:GaussCodazzi}
\eeq
where $\tilde{\IIFundForm}^\mu_{ab}$ is the traceless part of the second fundamental form
\beq
\tilde{\IIFundForm}^\mu_{ab} = \IIFundForm^\mu_{ab} - \frac{H^\mu}{2} h_{ab}\,.
\label{eqn:tracelessIIFundForm}
\eeq

\subsection{Embedding in normal coordinates}
\label{sec:NormalCoordinates}

Using these standard geometry results, we now derive the expressions needed
for~\eqref{eqn:FTexpansion} and~\eqref{eqn:DifferentialFormStructure}. Unlike
in Section~\ref{sec:abel}, we state here the result for a generic curved spacetime $M$. 
This allows us to perform the calculation in
Appendix~\ref{sec:GeodesicPointSplitting} on curved space.

Let $x^\mu$ and $\eta^a$ be Riemann 
normal coordinates on $M$ and $\Sigma$ about the same point. In terms of 
these, the embedding $\Sigma \hookrightarrow M$ may be expanded as
\beq
\label{EmbeddingNormalCoordinates}
x^\mu\left(\eta\right) = x^\mu(0) + \eta^a v_a^\mu + \frac{1}{2} \eta^a \eta^b v_{ab}^\mu 
+ \frac{1}{6} \eta^a \eta^b \eta^c v_{abc}^\mu + \ho[4]{\eta}\,.
\eeq
These coefficients are constrained by the condition that straight lines in
normal coordinates correspond to geodesics. In particular, a curve on $\Sigma$
given by a straight line in $\eta$ has constant speed and its curvature in $M$
is normal to $\Sigma$ at every point, which gives the constraints
\begin{equation}
  \begin{aligned}
  \delta_{ab} &= v_a \cdot v_b\,, \\
  0 &= v_{ab} \cdot v_c\,, \\
  0 &= 3 \: v_d \cdot v_{abc} + v_{ab} \cdot v_{cd} + v_{ac} \cdot v_{bd} + v_{ad}
  \cdot v_{bc}\,.
  \end{aligned}
\end{equation}
Using \eqref{eqn:IIFundForm} one easily checks that the second order coefficient equals the second 
fundamental form
\beq
\IIFundForm_{ab}^\mu\rvert_{\eta=0} = v_{ab}^\mu\,.
\eeq
The geodesic distance between $\xi(\eta)$ and the origin of the normal frame is 
found from \eqref{EmbeddingNormalCoordinates}
\beq
\label{GeodesicDistance}
\norm{x(\eta) - x(0)}^2 = \eta^a \eta_a - \frac{1}{12} \IIFundForm_{ab} \cdot
\IIFundForm_{cd} \eta^a \eta^b \eta^c \eta^d + \ho[5]{\eta}\,.
\eeq
Furthermore, in normal coordinates, the metrics take the form 
\beq
\label{MetricsNormalCoordinates}
\begin{split}
g_{\mu \nu} &= \delta_{\mu \nu} - \frac{1}{3} \RiemannTensor^M_{\mu \rho \nu
\sigma} \xi^\rho \xi^\sigma + \ho[3]{\xi}\,, \\
h_{ab} &= \delta_{ab} - \frac{1}{3} \RiemannTensor^\Sigma_{acbd} \eta^c \eta^d +
\ho[3]{\eta}\,,
\end{split}
\eeq
which yields an expansion for the volume factor 
\beq
\label{VolumeFactor}
\sqrt{h(\eta)} = 1 - \frac{1}{6} \RicciTensor^\Sigma_{ab} \eta^a \eta^b +
\ho[3]{\eta}\,.
\eeq

\section{Geodesic point-splitting}
\label{sec:GeodesicPointSplitting}

In this appendix we present an alternative regularisation
of~\eqref{I_Sigma_Definition}, essentially point splitting, displacing 
one copy of the surface operator by a distance $\epsilon$ in
an arbitrary normal direction $\nu$. This
regularisation is used in~\cite{gus03, gustavsson:2004gj}, but there the vector
$\nu$ is taken to be a constant, and therefore the method is only applicable if
the operators are restricted to a codimension-one subspace. 

The technology used to define this regularisation scheme applies for generic
smooth embedded surfaces in a Riemannian manifold, and we present here a curved
space calculation, as opposed to Section~\ref{sec:FTanomaly}, where for brevity
we restricted ourselves to flat space. However, we still have to restrict to
conformally flat backgrounds, since otherwise we do not have a short-distance
expansion for the propagator and therefore still cannot infer the anomaly
coefficient $b$.

As expected, we recover the result~\eqref{eqn:FTanomaly} exactly, and thus 
verify scheme-independence.

\subsection{Displacement map}
\label{sec:DisplacementMap}

We can regularise the integral~\eqref{I_Sigma_Definition} by
displacing a copy of the surface a
distance $\epsilon$ along a unit normal vector field $\nu$. Under that
map, which we denote by ${\cal T}$, 
the geodesic distance admits an expansion of the form
\beq
\label{eqn:GPSgeneralexpansion}
  \norm{{\cal T}(x^\mu(\sigma)) - x^\mu(\sigma + \eta)}^2 =
  \epsilon^2 + \eta^2 + 
  \sum_{k = 3}^\infty\sum_{l=0}^k f^{(k)}_l\eta^l\epsilon^{k-l}\,.
\eeq
We can combine the terms of fixed $k$ in terms of degree $k$ polynomials $f^{(k)}$
\beq
\sum_{l=0}^k f^{(k)}_l\eta^l\epsilon^{k-l}=\epsilon^k f^{(k)}(\eta/\epsilon)\,.
\eeq
We calculate the higher order terms in \eqref{eqn:GPSgeneralexpansion} explicitly
in~\eqref{DisplacementExpansionPointSplitting}, but first we note that the only
terms contributing to the divergent part are $f^{(3)}$ and $f^{(4)}$. To see that, 
the integrals computing the expectation value take the form
\beq
  \int\limits_0^\rho \frac{\eta^{m+1} \diff \eta}
  {\norm{{\cal T}(x^\mu(\sigma)) - x^\mu(\sigma + \eta)}^4}\,,
  \label{eqn:GPSregularisedintegrals}
\eeq
where $\rho$ is an arbitrary but fixed IR cutoff. We can
evaluate~\eqref{eqn:GPSregularisedintegrals} by expanding the integrand in $\epsilon$.
Writing $s \equiv \eta/\epsilon$, we obtain
\beq
  \label{eqn:GPSdivergences}
  \epsilon^{m-2} \int\limits_0^{\rho/\epsilon}
  \frac{s^{m+1}}{\left( 1 + s^2 \right)^2} \left[ 1
  - \frac{2 f^{(3)}(s)}{1+s^2} \epsilon 
  + \left( \frac{3 (f^{(3)}(s))^2}{(1 + s^2)^2}
  - \frac{2 f^{(4)}(s)}{1+s^2}\right) \epsilon^2 + \ho[3]{\epsilon}  \right] \diff s\,.
\eeq
By application of Fa\`a di Bruno's formula one checks that the terms in brackets of order $\epsilon^n$ 
contribute to the divergence only if $m+n \leq 2$. We can therefore safely 
ignore higher orders in $\epsilon$. Only a finite number of terms remains to be computed 
and we find that the only divergent integrals \eqref{eqn:GPSregularisedintegrals} are:
\begin{subequations}
  \begin{align}
  \label{eqn:GPSdivergences_m=0}
  m=0:&\quad
  \frac{1}{2\epsilon^2} 
  - \frac{1}{8\epsilon} \left( 4 f_0^{(3)} + \pi f^{(3)}_1 + 4 f^{(3)}_2 + 3 \pi f^{(3)}_3  \right)
  + \left( - 3 (f^{(3)}_3)^2 + 2 f^{(4)}_4 \right) \log \epsilon\,, \\
  \label{eqn:GPSdivergences_m=1}
  m=1:&\quad \frac{\pi}{4\epsilon} + 2 f^{(3)}_3 \log \epsilon\,, \\
  \label{eqn:GPSdivergences_m=2}
  m=2:&\quad {- \log \epsilon}\,.
  \end{align}
\end{subequations}

The relevant coefficients can be read off of the expansion of the geodesic distance 
up to combined order of 4 in $\eta$ and $\epsilon$. The second term on the 
left hand side of \eqref{eqn:GPSgeneralexpansion} can be expanded simply 
using the embedding \eqref{EmbeddingNormalCoordinates}. For the first term, 
we solve the geodesic equation order by order in the displacement $\epsilon$
to obtain
\beq
  {\cal T}(x^\mu) =  x^\mu +
  \epsilon \nu^\mu -
  \frac{\epsilon^2}{2}\Gamma^\mu_{\kappa\lambda} \nu^\kappa \nu^\lambda +
  \frac{\epsilon^3}{6} \left( 
  -\partial_\nu \Gamma^\mu_{\rho \sigma}
  + 2 \Gamma_{\nu\lambda}^\mu \Gamma_{\rho\sigma}^\lambda \right)
  \nu^\nu \nu^\rho \nu^\sigma +
  \ho[4]{\epsilon}\,.
  \label{eqn:displacement}
\eeq
Combining these expressions, and writing $\eta^a = \eta e^a(\varphi)$ as in
\eqref{eqn:FTpolaridentities} and onwards, 
the only two non-vanishing relevant coefficients read
\begin{equation}
  \begin{aligned}
  f^{(3)}_2 &= - e^a e^b \IIFundForm_{ab} \cdot \nu\,, \\
  f^{(4)}_4 &= - \frac{1}{12} e^a e^b e^c e^d \IIFundForm_{ab} \cdot \IIFundForm_{cd}\,.
  \label{DisplacementExpansionPointSplitting}
  \end{aligned}
\end{equation}
The first contributes to a scheme-dependent divergence $\epsilon^{-1}$, while
the second contributes to the anomaly.

\subsection{Evaluation of the anomaly}

With the displacement map~\eqref{eqn:displacement} in hand, we can
evaluate~\eqref{I_Sigma_Definition}. 
The propagators on a conformally flat 
background can be obtained by considering curved space actions for a conformal scalar 
and a Maxwell-type 2-form and inverting the kinetic operators order by order, 
following~\cite{gustavsson:2004gj} and~\cite{henningson:1999xi}. We find:
\begin{align}
\label{eqn:CurvedSpacePropagators}
  \vev{\Phi_i(x) \Phi_j(x+\xi)} &= \frac{\delta_{ij}}{2\pi^2 \norm{\xi}^4}
  \left[ 1 + \frac{1}{3} P_{\mu\nu} \xi^\mu \xi^\nu + \ho[3]{\xi} \right], \\
  \vev{B^{+\mu \nu}(x) B^+_{\rho \sigma}(x + \xi)} &=
  \frac{1}{2 \pi^2 \norm{\xi}^4}
  \Big[\delta_\mu^\rho\delta_\nu^\sigma-\delta_\nu^\rho\delta_\mu^\sigma
    \\&\qquad
    - \frac{4}{3} \left(
    4 P^{[\mu}_{[\rho} \delta^{\nu]}_{\sigma]} \delta_{\lambda \tau}
    + P_{\lambda [\rho} \delta^{[\mu}_{\sigma]} \delta_{\tau}^{\nu]} 
    + \delta_{\lambda [\rho} P^{[\mu}_{\sigma]} \delta_{\tau}^{\nu]} 
    \right) \xi^\lambda \xi^\tau + \ho[3]{\xi}\Big]
\,.
\nonumber
\end{align}
To apply our regularisation, we should replace $\xi$
by~\eqref{eqn:GPSgeneralexpansion} in the denominator of the propagators before
performing the integral over $\eta$. A priori, we should also perform the
displacement in the numerator, since a term of order $\ho{\epsilon}$ can
contribute to the $\epsilon^{-1}$ divergence by
multiplying~\eqref{eqn:GPSdivergences_m=0}. However, one easily checks that the
only terms of that order are accompanied by nonzero powers of $\eta$, and
therefore do not contribute to the divergence of~\eqref{I_Sigma_Definition}.  We
therefore drop the $\epsilon$ in the numerators of the propagators.

The expansion of the numerators is then assembled, as before, from 
\eqref{eqn:FTexpansion} and \eqref{eqn:DifferentialFormStructure},
but in addition, since we are working on curved space, we obtain an additional 
term at $\ho[2]{\eta} $ explicitly involving $\tr P$ from the 
propagators~\eqref{eqn:CurvedSpacePropagators}. 
Collecting terms in analogy to Section~\ref{sec:FTanomaly}, and integrating 
out the angular coordinate using~\eqref{eqn:FTpolaridentities}, we obtain the 
scalar contribution
\beq
\label{eqn:GPSscalarContribution}
\frac{1}{4\pi\epsilon^2} + \frac{H \cdot \nu }{8 \pi\epsilon} 
+ \frac{1}{32 \pi} \left( 2 \RicciScalar^\Sigma - \left( H^2 + 4 \tr P\right) 
+ 4 \cAnomaly \right) \log \epsilon + \fin,
\eeq
while the tensor field yields
\beq
\label{eqn:GPStensorContribution}
-\frac{1}{4 \pi\epsilon^2} - \frac{H \cdot \nu }{8 \pi\epsilon}
 - \frac{1}{32 \pi} \left( -2 \RicciScalar^\Sigma 
+ 3 \left(H^2 + 4 \tr P \right) \right) \log \epsilon + \fin.
\eeq
Combining these terms, we find 
\beq
\log \vev{V_\Sigma} = \frac{1}{8\pi} \log \epsilon \int_\Sigma \vol_\Sigma \left[ \RicciScalar^\Sigma 
- \left( H^2 + 4 \tr P \right) + \cAnomaly \right] + \fin,
\eeq
which agrees exactly with~\eqref{eqn:FTanomaly}. Note that the scheme dependence,
which is present in the simple pole of both~\eqref{eqn:GPSscalarContribution} 
and~\eqref{eqn:GPStensorContribution}, cancels in the final result, and the 
terms $H^2$ and $\tr P$ combine to an anomaly term as in~\eqref{eqn:SCAnomaly}, 
as required.

\bibliographystyle{utphys2}
\bibliography{ref}

\end{document}